\documentclass{article}
\usepackage{amsfonts}
\usepackage{amsmath}
\usepackage{cite}

\input{tcilatex}
\begin{document}

\title{Path integral discussion of the improved Tietz potential}
\author{A. Khodja, F. Benamira and L. Guechi \\
Laboratoire de Physique Th\'{e}orique, D\'{e}partement de Physique, \and %
Facult\'{e} des Sciences Exactes, Universit\'{e} des fr\`{e}res Mentouri,
\and Route d'Ain El Bey, Constantine, Algeria}
\maketitle

\begin{abstract}
An improved form of the Tietz potential for diatomic molecules is \
discussed in detail within the path integral formalism. The radial Green's
function is rigorously constructed in a closed form for different shapes of
this potential. For $\left\vert q\right\vert \leq 1,$ and $\frac{1}{2\alpha }%
\ln \left\vert q\right\vert <r<+\infty $, the energy spectrum and the
normalized wave functions of the bound states are derived for the $l$ waves.
When the deformation parameter $q$ is $0<\left\vert q\right\vert <1$ or $q>0$%
, it is found that the quantization conditions are transcendental equations
that requires numerical solutions. In the limit $q\rightarrow 0$, the energy
spectrum and the corresponding wave functions for the radial Morse potential
are recovered.


03.65.Db-Functional analytical methods

\end{abstract}

\section{Introduction}

As an advantageous alternative to the Morse potential for fitting certain
diatomic interactions potential energy curves, Tietz \cite{Tietz} suggested,
in 1963, a potential energy function given by

\begin{equation}
U(r)=D_{e}\left[ 1+\frac{\left( a+b\right) e^{-2\beta r}-be^{-\beta r}}{%
\left( 1+ce^{-\beta r}\right) ^{2}}\right]  \label{F1a}
\end{equation}%
which continues to stimulate numerous theoretical works \cite%
{Wei,Natanson,Kunc,Gordillo,Hamzavi1,Hamzavi2,Ikhdair,Ikot,Roy,Falaye1,
Falaye2,Guechi} made within the framework of various algebraic methods. In
2012, Jia et al \cite{Jia1} proposed an improved version of this potential
having the form:

\begin{equation}
\text{\ }V_{T}(r)=D_{e}\left( 1-\frac{e^{2\alpha r_{e}}+q}{e^{2\alpha r}+q}%
\right) ^{2},\text{\ \ }  \label{F1b}
\end{equation}%
in which $D_{e}$ denotes the dissociation energy of the diatomic molecule, $%
r_{e}$ is the equilibrium bond length, $\alpha ^{-1}$ is the range of the
potential and $\ q$ is a real deformation parameter. The introduction of the
parameter $q$ can serve as an additional parameter in describing
inter-atomic interactions, and especially in three-dimensional problems, it
allows to establish the center of mass location of a molecule at a certain
distance from the coordinate origin. This potential has received a
considerable attention immediately afterwards. The equivalence between the
deformed Rosen-Morse potential and the Tietz potential for diatomic
molecules has been studied \cite{Chen,Sun}. Note also that the improved
Tietz potential and the modified Rosen-Morse potential have been
investigated from different points of view in these last years \cite%
{Liu,Hu,Tang,Zhang,Dai}. Another interesting case, which despite
appearances, presents some kinship with the potentials above is the new
deformed Schi\"{o}berg-type potential introduced by Mustafa \cite{Mustafa}
to calculate the ro-vibrational energy levels of some diatomic molecules in
the context of the supersymmetric quantum mechanics. This potential has also
been discussed in an approach based on the Feynman path integral \cite%
{Amrouche}.

The potential function (\ref{F1b}) contains three kinds of potentials namely
the deformed modified Manning-Rosen potential \cite{Manning} for $q<0$, the
deformed modified Rosen-Morse potential \cite{Rosen} when $q>0$ and the
Morse potential \cite{Morse} at the limit $q\rightarrow 0$. Therefore, it is
clear that the parameter $q$ will have an influence on the form of the
solutions of the problem. For $q\neq 0$, we will consider each case
separately using the path integral method.

Our paper is organized as follows. In section 2, we shortly formulate the
path integral for the Green's function associated with the spherically
symmetric potential $V_{T}(r).$ In section 3, we examine the problem for $%
q<0 $ by distinguishing two cases: $q\leq -1$ and $-1<q<0$. When $q\leq -1$
and $\frac{1}{2\alpha }\ln \left\vert q\right\vert <r<+\infty $, we
calculate the radial Green's function associated with the deformed modified
Manning-Rosen potential\ for a state of orbital momentum $l$ by using a
proper approximation to the centrifugal potential term to find the energy
spectrum and the normalized wave functions. For \ $-1<q<0$, the $\left\vert
q\right\vert -$deformed Manning-Rosen potential is converted into the
standard Manning-Rosen potential defined in the half-space $\xi >\frac{1}{%
2\alpha }\ln \left\vert q\right\vert $. Following the procedure used in our
earlier works \cite{Benamira1,Benamira2,Khodja}, it is then fairly simple to
write down the radial Green's function for the $s-$waves in a closed form
and deduce from it a transcendental equation for the energy levels and the
non-normalized wave functions. In section 4, the potential (\ref{F1b}), for $%
q>0$, is worked out in a similar way. We first transform it into a $q-$%
deformed modified Rosen-Morse potential and consequently into the standard
Rosen-Morse potential defined in the half-space $u>\frac{1}{2\alpha }\ln q$.
We construct the radial Green's function in a closed form from which we also
obtain a transcendental equation for the $s-$state energy levels and the
non-normalized wave functions. In the limit $q\rightarrow 0$, we recover the
results of the radial Morse potential problem for the Green's function,
energy spectrum and wave functions in section 5. In section 6, we briefly
discuss some special cases and compare our results with those given in the
literature by other authors. Section 7 will be a conclusion.

\section{Green's function}

The energy-dependent Green's function for a particle of mass $M$, moving in
the potential (\ref{F1b}) can be expanded into partial waves \cite{Peak} in
spherical coordinates

\begin{equation}
G(\mathbf{r}^{\prime \prime },\mathbf{r}^{\prime };E)=\frac{1}{r^{\prime
\prime }r^{\prime }}\overset{\infty }{\underset{l=0}{\sum }}\frac{2l+1}{4\pi 
}G_{l}(r^{\prime \prime },r^{\prime };E)P_{l}(\cos \Theta ),  \label{F2}
\end{equation}%
in which the radial Green's function is given by

\begin{equation}
G_{l}(r^{\prime \prime },r^{\prime };E)=\int_{0}^{\infty }dT\exp \left( 
\frac{i}{\hbar }ET\right) \left\langle r^{\prime \prime }\right\vert \exp
\left( -\frac{i}{\hbar }\widehat{H}_{l}T\right) \left\vert r^{\prime
}\right\rangle ,  \label{F3}
\end{equation}%
and $P_{l}(\cos \Theta )$ is the Legendre polynomial of degree $l$ in $\cos
\Theta =\frac{\overrightarrow{r}^{\prime \prime }\overrightarrow{r}^{\prime }%
}{r^{\prime \prime }r^{\prime }}=\cos \theta ^{\prime \prime }\cos \theta
^{\prime }+\sin \theta ^{\prime \prime }\sin \theta ^{\prime }\cos (\phi
^{\prime \prime }-\phi ^{\prime })$. The Hamiltonian operator $\widehat{H}%
_{l}$ is defined by 
\begin{equation}
\widehat{H}_{l}=\frac{\widehat{P}_{r}^{2}}{2M}+\frac{\hbar ^{2}l(l+1)}{%
2Mr^{2}}+V_{T}(r),  \label{F4}
\end{equation}%
where $\widehat{P}_{r}=-i\hbar (\partial /r\partial r)r$ is the radial
momentum operator.

In the path integral formalism, the integrand $\left\langle r^{\prime \prime
}\right\vert \exp \left( -\frac{i}{\hbar }\widehat{H}_{l}T\right) \left\vert
r^{\prime }\right\rangle $ is explicitly given by the Hamiltonian path
integral,%
\begin{eqnarray}
\left\langle r^{\prime \prime }\right\vert \exp \left( -\frac{i}{\hbar }%
\widehat{H}_{l}T\right) \left\vert r^{\prime }\right\rangle &=&\underset{%
N\rightarrow \infty }{\lim }\overset{N}{\underset{n=1}{\dprod }}\left[ \int
dr_{n}\right] \overset{N+1}{\underset{n=1}{\dprod }}\left[ \int \frac{%
d\left( P_{r}\right) _{n}}{2\pi \hbar }\right]  \notag \\
&&\times \exp \left\{ \frac{i}{\hbar }\overset{N+1}{\underset{n=1}{\sum }}%
\mathcal{A}_{1}^{n}\right\} ,  \label{F5}
\end{eqnarray}%
where the short-time action is

\begin{equation}
\mathcal{A}_{1}^{n}=\left( P_{r}\right) _{n}\Delta r_{n}-\varepsilon \left[ 
\frac{\left( P_{r}\right) _{n}^{2}}{2M}+\frac{\hbar ^{2}l(l+1)}{2Mr_{n}^{2}}%
+V_{T}(r_{n})\right] .  \label{F6}
\end{equation}%
Note that $\left( P_{r}\right) _{n}=P_{r}(t_{n}),$ $r_{n}=r(t_{n}),$ $\Delta
r_{n}=r_{n}-r_{n-1},$ $\varepsilon =t_{n}-t_{n-1}$ and $T=\left( N+1\right)
\varepsilon .$ The momentum integrations can be carried out and we obtain
the Lagrangian path integral,%
\begin{eqnarray}
\left\langle r^{\prime \prime }\right\vert \exp \left( -\frac{i}{\hbar }%
\widehat{H}_{l}T\right) \left\vert r^{\prime }\right\rangle &=&\underset{%
N\rightarrow \infty }{\lim }\left( \frac{M}{2i\pi \hbar \varepsilon }\right)
^{\frac{N+1}{2}}\overset{N}{\underset{n=1}{\dprod }}\left[ \int dr_{n}\right]
\notag \\
&&\times \exp \left\{ \frac{i}{\hbar }\overset{N+1}{\underset{n=1}{\sum }}%
\left[ \frac{M}{2\varepsilon }\left( \Delta r_{n}\right) ^{2}-\varepsilon
\left( \frac{\hbar ^{2}l(l+1)}{2Mr_{n}^{2}}+V_{T}(r_{n})\right) \right]
\right\} .  \notag \\
&&  \label{F7}
\end{eqnarray}%
To construct the radial Green's function for $V_{T}(r)$ via (\ref{F3}) in
terms of the path integral formalism \cite{Kleinert}, three cases may be
distinguished according to the real values of the parameter $q$.

\section{Deformed modified Manning-Rosen potential}

When the deformation parameter $q$ is negative, the improved Tietz potential
is equivalent to the deformed modified Manning-Rosen potential

\begin{equation}
V_{\left\vert q\right\vert }(r)=U_{0}-U_{1}\coth _{\left\vert q\right\vert
}\left( \alpha r\right) +U_{2}\coth _{\left\vert q\right\vert }^{2}\left(
\alpha r\right) ,  \label{F8}
\end{equation}%
where $U_{0},$ $U_{1}$ and $U_{2}$ are the real constants given by 
\begin{equation}
\left\{ 
\begin{array}{c}
U_{0}=\frac{D_{e}}{4}\left( \frac{e^{2\alpha r_{e}}}{_{\left\vert
q\right\vert }}+1\right) ^{2}, \\ 
U_{1}=\frac{D_{e}}{2}\left( \frac{e^{2\alpha r_{e}}}{_{\left\vert
q\right\vert }}+1\right) \left( \frac{e^{2\alpha r_{e}}}{_{\left\vert
q\right\vert }}-1\right) , \\ 
U_{2}=\frac{D_{e}}{4}\left( \frac{e^{2\alpha r_{e}}}{_{\left\vert
q\right\vert }}-1\right) ^{2}.%
\end{array}%
\right.  \label{F9}
\end{equation}%
The potential (\ref{F8}) is obtained by using a $q-$deformation of the usual
hyperbolic functions denoted by

\begin{equation}
\sinh _{q}x=\frac{e^{x}-qe^{-x}}{2},\text{ }\cosh _{q}x=\frac{e^{x}+qe^{-x}}{%
2},\text{ }\tanh _{q}x=\frac{\sinh _{q}x}{\cosh _{q}x},\text{ }\coth _{q}x=%
\frac{\cosh _{q}x}{\sinh _{q}x}.  \label{F10}
\end{equation}%
These functions have been introduced for first time by Arai \cite{Arai} ,
with the real parameter\ $q>0$. The deformation parameter $q$ can be
extended to the cases of $q<0$ and the complex number \cite{Jia2} .

Now, the energy-dependent Green's for the deformed modified Manning-Rosen
potential is given by

\begin{equation}
G_{l}^{\left\vert q\right\vert }(r^{\prime \prime },r^{\prime
};E)=\int_{0}^{\infty }dT\exp \left( \frac{i}{\hbar }ET\right)
K_{l}^{\left\vert q\right\vert }\left( r^{\prime \prime },r^{\prime
};T\right) ,  \label{F11}
\end{equation}%
where the propagator $K_{l}^{\left\vert q\right\vert }\left( r^{\prime
\prime },r^{\prime };T\right) $ is formally written as: 
\begin{equation}
K_{l}^{\left\vert q\right\vert }\left( r^{\prime \prime },r^{\prime
};T\right) =\int \mathit{D}r(t)\exp \left\{ \frac{i}{\hbar }\int_{0}^{T}%
\left[ \frac{M}{2}\overset{.}{r}^{2}-\left( \frac{\hbar ^{2}l(l+1)}{2Mr^{2}}%
+V_{\left\vert q\right\vert }(r)\right) \right] dt\right\} .  \label{F12}
\end{equation}%
In order to evaluate (\ref{F11}), we have to distinguish two cases: $%
\left\vert q\right\vert \geq 1$ and $0<\left\vert q\right\vert <1$.

\subsection{First case: $\left\vert q\right\vert \geq 1$ and $r_{0}<r<\infty 
$}

When $\left\vert q\right\vert \geq 1$, the potential (\ref{F8}) has a strong
singularity at the point $r_{0}=\frac{1}{2\alpha }\ln \left\vert
q\right\vert $, creating an impenetrable barrier. In this case, there are
two distinct regions, one is defined by the interval $\left] 0,r_{0}\right[ $
and the other by the interval $\left] r_{0},\infty \right[ $. As in the
first interval, the calculation of the Green's function is without physical
interest, we will construct the latter only in the second interval for a
state of orbital momentum $l$ by using the following approximation to deal
with the centrifugal potential term \cite{Jia3}:%
\begin{equation}
\frac{1}{r^{2}}\approx C_{0}+\frac{B_{0}}{e^{2\alpha r}-\left\vert
q\right\vert }+\frac{A_{0}}{\left( e^{2\alpha r}-\left\vert q\right\vert
\right) ^{2}},\text{ for }\alpha r\ll 1\text{ and }\left\vert q\right\vert
\geq 1\text{,}  \label{F13}
\end{equation}%
where $C_{0}=\frac{\alpha ^{2}}{12},B_{0}$ and $A_{0}$ are two ajustable
parameters. If we take $C_{0}=0,B_{0}=\frac{A_{0}}{\left\vert q\right\vert }%
,A_{0}=\alpha ^{2}\left\vert q\right\vert ^{2}$ and $\left\vert q\right\vert
=1,$ Eq. (\ref{F13}) is reduced to the approximation proposed by Greene and
Aldrich \cite{Greene}.

With (\ref{F13}) and (\ref{F10}), the effective potential is written as:%
\begin{eqnarray}
V_{eff}(r) &=&\frac{\hbar ^{2}l(l+1)}{2Mr^{2}}+V_{\left\vert q\right\vert
}(r)  \notag \\
&\approx &V_{0}^{l}-V_{1}^{l}\coth _{\left\vert q\right\vert }\left( \alpha
r\right) +\frac{V_{2}^{l}}{\sinh _{\left\vert q\right\vert }^{2}\left(
\alpha r\right) },  \label{F14}
\end{eqnarray}%
where%
\begin{equation}
\left\{ 
\begin{array}{c}
V_{0}^{l}=\frac{\hbar ^{2}l(l+1)}{2M}\left[ C_{0}+\frac{1}{2\left\vert
q\right\vert }\left( \frac{A_{0}}{\left\vert q\right\vert }-B_{0}\right) %
\right] +U_{0}+U_{2}, \\ 
V_{1}^{l}=\frac{\hbar ^{2}l(l+1)}{4M\left\vert q\right\vert }\left( \frac{%
A_{0}}{\left\vert q\right\vert }-B_{0}\right) +U_{1}, \\ 
V_{2}^{l}=\frac{\hbar ^{2}l(l+1)A_{0}}{8M\left\vert q\right\vert }%
+\left\vert q\right\vert U_{2}.%
\end{array}%
\right.  \label{F15}
\end{equation}%
Then, changing $\alpha r$ into $\xi =\alpha r-\frac{1}{2}\ln \left\vert
q\right\vert $ and $\varepsilon $ into $\alpha ^{-2}\varepsilon _{s}$, we
can rewrite the Green's function (\ref{F11}) in the form%
\begin{equation}
G_{l}^{\left\vert q\right\vert \geq 1}(r^{\prime \prime },r^{\prime };E)=%
\frac{1}{\alpha }G_{MR}^{\left\vert q\right\vert \geq 1}(\xi ^{\prime \prime
},\xi ^{\prime };\widetilde{E}_{l}),  \label{F16}
\end{equation}%
where 
\begin{equation}
G_{MR}^{\left\vert q\right\vert \geq 1}(\xi ^{\prime \prime },\xi ^{\prime };%
\widetilde{E}_{l})=\int_{0}^{\infty }dS\exp \left( \frac{i}{\hbar }\frac{%
\widetilde{E}_{l}}{\alpha ^{2}}S\right) P_{MR}^{l}\left( \xi ^{\prime \prime
},\xi ^{\prime };S\right)  \label{F17}
\end{equation}%
with%
\begin{equation}
\widetilde{E}_{l}=E-\left( U_{0}+U_{2}+\frac{\hbar ^{2}l(l+1)}{2M}\left[
C_{0}+\frac{1}{2\left\vert q\right\vert }\left( \frac{A_{0}}{\left\vert
q\right\vert }-B_{0}\right) \right] \right) ,  \label{F18}
\end{equation}%
and%
\begin{equation}
P_{MR}^{l}\left( \xi ^{\prime \prime },\xi ^{\prime };S\right) =\int \mathit{%
D}\xi (s)\exp \left\{ \frac{i}{\hbar }\int_{0}^{S}\left[ \frac{M}{2}\overset{%
.}{\xi }^{2}-V_{MR}^{l}(\xi )\right] ds\right\}  \label{F19}
\end{equation}%
is the propagator for the standard Manning-Rosen potential \cite{Manning}%
\begin{equation}
V_{MR}^{l}(\xi )=-\frac{V_{1}^{l}}{\alpha ^{2}}\coth \xi +\frac{V_{2}^{l}}{%
\alpha ^{2}\left\vert q\right\vert \sinh ^{2}\xi }\text{; \ }\xi >0\text{,}
\label{F20}
\end{equation}%
which has been discussed in the literature by means of the path integral 
\cite{Grosche}. We can thus write down the solution of (\ref{F17})
immediately in a closed form as:%
\begin{eqnarray}
G_{MR}^{\left\vert q\right\vert \geq 1}(\xi ^{\prime \prime },\xi ^{\prime };%
\widetilde{E}_{l}) &=&-\frac{iM}{\hbar }\frac{\Gamma \left(
M_{1}-L_{E}\right) \Gamma \left( L_{E}+M_{1}+1\right) }{\Gamma \left(
M_{1}+M_{2}+1\right) \Gamma (M_{1}-M_{2}+1)}  \notag \\
&&\times \left( \frac{2}{\coth \xi ^{\prime }+1}\frac{2}{\coth \xi ^{\prime
\prime }+1}\right) ^{\frac{M_{1}+M_{2}+1}{2}}  \notag \\
&&\times \left( \frac{\coth \xi ^{\prime }-1}{\coth \xi ^{\prime }+1}\frac{%
\coth \xi ^{\prime \prime }-1}{\coth \xi ^{\prime \prime }+1}\right) ^{\frac{%
M_{1}-M_{2}}{2}}  \notag \\
&&\times \text{ }_{2}F_{1}\left( M_{1}-L_{E},L_{E}+M_{1}+1,M_{1}-M_{2}+1;%
\frac{\coth \xi _{>}-1}{\coth \xi _{>}+1}\right)  \notag \\
&&\times \text{ }_{2}F_{1}\left( M_{1}-L_{E},L_{E}+M_{1}+1,M_{1}+M_{2}+1;%
\frac{2}{\coth \xi _{<}+1}\right) ,  \notag \\
&&  \label{F21}
\end{eqnarray}%
where the symbols $\xi _{>}$ and $\xi _{<}$ denote$\ \max (\xi ^{\prime
\prime },\xi ^{\prime })$ and $\min (\xi ^{\prime \prime },\xi ^{\prime })$
respectively. $_{2}F_{1}(\alpha ,\beta ,\gamma ;z)$ is the hypergeometric
function and the quantities $L_{E},$ $M_{1}$ and $M_{2}$ are defined by 
\begin{equation}
\left\{ 
\begin{array}{c}
L_{E}=-\frac{1}{2}+\frac{1}{2\alpha }\sqrt{\frac{2M}{\hbar ^{2}}\left(
U_{0}+U_{1}+U_{2}-E\right) +l(l+1)\left( C_{0}+\frac{A_{0}}{\left\vert
q\right\vert ^{2}}-\frac{B_{0}}{\left\vert q\right\vert }\right) }, \\ 
M_{1}=\delta _{l}+\frac{1}{2\alpha }\sqrt{\frac{2M}{\hbar ^{2}}\left(
U_{0}-U_{1}+U_{2}-E\right) +l(l+1)C_{0}}, \\ 
M_{2}=\delta _{l}-\frac{1}{2\alpha }\sqrt{\frac{2M}{\hbar ^{2}}\left(
U_{0}-U_{1}+U_{2}-E\right) +l(l+1)C_{0}},%
\end{array}%
\right.  \label{F22}
\end{equation}%
with%
\begin{equation}
\delta _{l}=\frac{1}{2}\sqrt{1+\frac{8MU_{2}}{\hbar ^{2}\alpha ^{2}}+\frac{%
l(l+1)}{\alpha ^{2}}\frac{A_{0}}{\left\vert q\right\vert ^{2}}}.  \label{F23}
\end{equation}%
Substituting (\ref{F21}) into (\ref{F16}) and going back to the old variable
we get 
\begin{eqnarray}
G_{l}^{\left\vert q\right\vert \geq 1}(r^{\prime \prime },r^{\prime };E) &=&-%
\frac{iM}{\hbar \alpha }\frac{\Gamma \left( M_{1}-L_{E}\right) \Gamma \left(
L_{E}+M_{1}+1\right) }{\Gamma \left( M_{1}+M_{2}+1\right) \Gamma
(M_{1}-M_{2}+1)}  \notag \\
&&\times \left[ \left( 1-\left\vert q\right\vert e^{-2\alpha r^{\prime
}}\right) \left( 1-\left\vert q\right\vert e^{-2\alpha r^{\prime \prime
}}\right) \right] ^{\frac{M_{1}+M_{2}+1}{2}}  \notag \\
&&\times \left[ \left\vert q\right\vert ^{2}e^{-2\alpha (r^{\prime \prime
}+r^{\prime })}\right] ^{\frac{M_{1}-M_{2}}{2}}  \notag \\
&&\times \text{ }_{2}F_{1}\left(
M_{1}-L_{E},L_{E}+M_{1}+1,M_{1}-M_{2}+1;\left\vert q\right\vert e^{-2\alpha
r_{>}}\right)  \notag \\
&&\times \text{ }_{2}F_{1}\left(
M_{1}-L_{E},L_{E}+M_{1}+1,M_{1}+M_{2}+1;1-\left\vert q\right\vert
e^{-2\alpha r_{<}}\right) .  \notag \\
&&  \label{F24}
\end{eqnarray}%
The poles of the radial Green's function (\ref{F24}) in the complex
energy-plane corresponding to the bound states are all contained in the
first gamma function in the numerator. From the condition%
\begin{equation}
M_{1}-L_{E}=-n_{r},\text{ \ \ }n_{r}=0,1,2,...\text{ ,}  \label{F25}
\end{equation}%
and after inserting the expressions of $L_{E}$ and $M_{1}$ in (\ref{F22}),
we find the following expression for the vibrational and the rotational
energy levels of the diatomic molecule that has the value $l$ of the orbital
quantum number, and $n_{r}$ of the radial quantum number:%
\begin{eqnarray}
E_{n_{r},l}^{\left\vert q\right\vert \geq 1} &=&U_{0}+U_{2}+\frac{\hbar
^{2}l(l+1)}{2M}\left( \frac{A_{0}}{2\left\vert q\right\vert ^{2}}-\frac{B_{0}%
}{2\left\vert q\right\vert }+C_{0}\right) -\frac{\hbar ^{2}\alpha ^{2}}{2M}%
\left( N_{r}^{2}+\frac{\lambda _{l}^{2}}{N_{r}^{2}}\right) ,  \notag \\
&&  \label{F26}
\end{eqnarray}%
where%
\begin{equation}
N_{r}=n_{r}+\delta _{l}+\frac{1}{2},  \label{F27}
\end{equation}%
and%
\begin{equation}
\lambda _{l}=\frac{MU_{1}}{\hbar ^{2}\alpha ^{2}}+\frac{l(l+1)}{4\alpha ^{2}}%
\left( \frac{A_{0}}{\left\vert q\right\vert ^{2}}-\frac{B_{0}}{\left\vert
q\right\vert }\right) .  \label{F28}
\end{equation}

To obtain the reduced radial wave functions, we approximate the gamma
function $\Gamma \left( M_{1}-L_{E}\right) $ near the poles $%
M_{1}-L_{E}=-n_{r}$ as follows:%
\begin{eqnarray}
\Gamma \left( M_{1}-L_{E}\right) &\approx &\frac{\left( -1\right) ^{n_{r}}}{%
n_{r}!}\frac{1}{M_{1}-L_{E}+n_{r}}  \notag \\
&&=\frac{\left( -1\right) ^{n_{r}+1}}{n_{r}!}\frac{\hbar ^{2}\alpha ^{2}}{%
N_{r}M}\frac{\left( \frac{\lambda _{l}}{N_{r}}+N_{r}\right) \left( \frac{%
\lambda _{l}}{N_{r}}-N_{r}\right) }{E-E_{n_{r},l}^{\left\vert q\right\vert
\geq 1}},  \label{F29}
\end{eqnarray}%
and take into account the Gauss's transformation formula (see Ref. \cite%
{Gradshtein}, p. 1043, Eq. (9.131.2))%
\begin{eqnarray}
_{2}F_{1}(a,b,c;z) &=&\frac{\Gamma (c)\Gamma (c-a-b)}{\Gamma (c-a)\Gamma
(c-b)}\text{ }_{2}F_{1}(a,b,a+b-c+1;1-z)  \notag \\
&&+\frac{\Gamma (c)\Gamma (a+b-c)}{\Gamma (a)\Gamma (b)}(1-z)^{c-a-b}\text{ }%
_{2}F_{1}(c-a,c-b,c-a-b+1;1-z),  \notag \\
&&  \label{F30}
\end{eqnarray}%
in which, in this case, the second term is null because the gamma function $%
\Gamma (a)$ is infinite $(a=M_{1}-L_{E}=-n_{r}\leq 0).$ Thus, we arrive at
an expression in the form of a spectral expansion for the radial Green's
function (\ref{F24}),%
\begin{equation}
G_{l}^{\left\vert q\right\vert \geq 1}(r^{\prime \prime },r^{\prime
};E)=i\hbar \underset{n_{r}=0}{\overset{n_{r\max }}{\sum }}\frac{\chi
_{n_{r},l}^{\left\vert q\right\vert \geq 1}(r^{\prime \prime })\chi
_{n_{r},l}^{\left\vert q\right\vert \geq 1}(r^{\prime })}{%
E-E_{n_{r,l}}^{\left\vert q\right\vert \geq 1}},  \label{F31}
\end{equation}%
where the reduced (normalized) radial wave functions are given by%
\begin{eqnarray}
\chi _{n_{r},l}^{\left\vert q\right\vert \geq 1}(r) &=&C_{n_{r},l}\left(
1-\left\vert q\right\vert e^{-2\alpha r}\right) ^{\delta _{l}+\frac{1}{2}%
}\left( \left\vert q\right\vert e^{-2\alpha r}\right) ^{\frac{1}{2}\left( 
\frac{\lambda _{l}}{N_{r}}-N_{r}\right) }  \notag \\
&&\times \text{ }_{2}F_{1}\left( -n_{r},\frac{\lambda _{l}}{N_{r}}%
+N_{r}-n_{r},\frac{\lambda _{l}}{N_{r}}-N_{r}+1;\left\vert q\right\vert
e^{-2\alpha r}\right) ,  \label{F32}
\end{eqnarray}%
with the correct normalization factor 
\begin{eqnarray}
C_{n_{r},l} &=&\left[ \frac{\alpha }{N_{r}}\frac{\left( \frac{\lambda _{l}}{%
N_{r}}+N_{r}\right) \left( \frac{\lambda _{l}}{N_{r}}-N_{r}\right) \Gamma
\left( \frac{\lambda _{l}}{N_{r}}+N_{r}-n_{r}\right) \Gamma \left( 1+n_{r}+%
\frac{\lambda _{l}}{N_{r}}-N_{r}\right) }{n_{r}!\Gamma \left(
2N_{r}-n_{r}\right) \Gamma ^{2}\left( 1+\frac{\lambda _{l}}{N_{r}}%
-N_{r}\right) }\right] ^{\frac{1}{2}}.\,\,  \notag \\
&&  \label{F33}
\end{eqnarray}

Using the connecting formula (see Ref. \cite{Gradshtein}, p.952, Eq.
(8.406.1))%
\begin{equation}
P_{n}^{\left( \alpha ,\beta \right) }\left( t\right) =\frac{\Gamma \left(
n+\alpha +1\right) }{n!\Gamma \left( \alpha +1\right) }\text{ }%
_{2}F_{1}\left( -n,n+\alpha +\beta +1,\alpha +1;\frac{1-t}{2}\right) ,
\label{F34}
\end{equation}%
between the hypergeometric function and the Jacobi polynomial $P_{n}^{\left(
\alpha ,\beta \right) }\left( t\right) $, we can express (\ref{F32}) in the
form%
\begin{eqnarray}
\chi _{n_{r},l}^{\left\vert q\right\vert \geq 1}(r) &=&\left[ \frac{\alpha }{%
N_{r}}\frac{\left( \frac{\lambda _{l}}{N_{r}}+N_{r}\right) \left( \frac{%
\lambda _{l}}{N_{r}}-N_{r}\right) n_{r}!\Gamma \left( \frac{\lambda _{l}}{%
N_{r}}+N_{r}-n_{r}\right) }{\Gamma \left( 2N_{r}-n_{r}\right) \Gamma \left(
1+n_{r}+\frac{\lambda _{l}}{N_{r}}-N_{r}\right) }\right] ^{\frac{1}{2}} 
\notag \\
&&\times \left( \left\vert q\right\vert e^{-2\alpha r}\right) ^{\frac{1}{2}%
\left( \frac{\lambda _{l}}{N_{r}}-N_{r}\right) }\left( 1-\left\vert
q\right\vert e^{-2\alpha r}\right) ^{\delta _{l}+\frac{1}{2}}  \notag \\
&&P_{n_{r}}^{\left( \frac{\lambda _{l}}{N_{r}}-N_{r},2\delta _{l}\right)
}\left( 1-2\left\vert q\right\vert e^{-2\alpha r}\right) .  \label{F35}
\end{eqnarray}%
Now, for them to be physically acceptable, the wave functions $\chi
_{n_{r},l}^{\left\vert q\right\vert \geq 1}(r)$ must satisfy the boundary
conditions%
\begin{equation}
\underset{r\rightarrow r_{0}}{\lim }\chi _{n_{r},l}^{\left\vert q\right\vert
\geq 1}(r)=0,  \label{F36}
\end{equation}%
and%
\begin{equation}
\underset{r\rightarrow \infty }{\lim }\chi _{n_{r},l}^{\left\vert
q\right\vert \geq 1}(r)=0.  \label{F37}
\end{equation}%
When $r\rightarrow r_{0}$, it is obvious that (\ref{F35}) fulfills the
boundary condition (\ref{F36}), but, by letting $r\rightarrow \infty $ in (%
\ref{F35}), we obtain the asymptotic behavior%
\begin{equation}
\chi _{n_{r},l}^{\left\vert q\right\vert \geq 1}(r)\underset{r\rightarrow
\infty }{\sim }\left( e^{-\alpha r}\right) ^{\left( \frac{\lambda _{l}}{N_{r}%
}-N_{r}\right) },  \label{F38}
\end{equation}%
from which we must impose the restriction that only the wave functions with $%
\left( \frac{\lambda _{l}}{N_{r}}-N_{r}\right) >0$ fulfill the boundary
condition (\ref{F37}) and therefore the value of ~$n_{r\max }$ in (\ref{F31}%
) is given by $n_{r\max }=\left\{ \sqrt{\lambda _{l}}-\delta _{l}-\frac{1}{2}%
\right\} $ which denotes the largest integer $n_{r}\in 
\mathbb{N}
$ with $n_{r}<$ $\left( \sqrt{\lambda _{l}}-\delta _{l}-\frac{1}{2}\right) .$%
This condition provides a finite number of energy levels of the physical
system.

\subsection{Second case: $0<\left\vert q\right\vert <1$ and $r\in 
\mathbb{R}
^{+}$}

In this case, we limit ourselves to the evaluation of the Green's function
associated with the $s-$waves $(l=0)$. Making the change of variable defined
by $\xi =\alpha r-\frac{1}{2}\ln \left\vert q\right\vert $, $\left( \xi \in %
\left] \xi _{0},\infty \right[ \text{, }\xi _{0}=-\frac{1}{2}\ln \left\vert
q\right\vert \right) $ and performing the time transformation $\frac{dt}{ds}=%
\frac{1}{\alpha ^{2}}$, we can rewrite (\ref{F11}), for $l=0$, as%
\begin{eqnarray}
G_{0}^{0<\left\vert q\right\vert <1}(r^{\prime \prime },r^{\prime };E) &=&%
\frac{1}{\alpha }\widetilde{G}(\xi ^{\prime \prime },\xi ^{\prime };%
\widetilde{E}_{0})  \notag \\
&=&\frac{1}{\alpha }\int_{0}^{\infty }dS\exp \left( \frac{i}{\hbar }%
\widetilde{E}_{0}S\right) K_{0}^{0<\left\vert q\right\vert <1}(\xi ^{\prime
\prime },\xi ^{\prime };S),  \label{F39}
\end{eqnarray}%
where%
\begin{equation}
\widetilde{E}_{0}=\frac{1}{\alpha ^{2}}\left[ E-\left( U_{0}+U_{2}\right) %
\right] ,  \label{F40}
\end{equation}%
and%
\begin{equation}
K_{0}^{0<\left\vert q\right\vert <1}(\xi ^{\prime \prime },\xi ^{\prime
};S)=\int \mathit{D}\xi (s)\exp \left\{ \frac{i}{\hbar }\int_{0}^{S}\left( 
\frac{M}{2}\overset{.}{\xi }^{2}-\widetilde{V}(\xi )\right) ds\right\} ,
\label{F41}
\end{equation}%
with%
\begin{equation}
\widetilde{V}(\xi )=-\frac{U_{1}}{\alpha ^{2}}\coth \xi +\frac{U_{2}}{\alpha
^{2}\sinh ^{2}\xi };\text{ \ \ }\xi >\xi _{0}.  \label{F42}
\end{equation}%
Note that the expression (\ref{F42}) is that of the Manning-Rosen potential 
\cite{Manning} in the range $\xi >\xi _{0}.$ This means that the kernel $%
K_{0}^{0<\left\vert q\right\vert <1}(\xi ^{\prime \prime },\xi ^{\prime };S)$
is the propagator describing the motion of a particle subjected to the
Manning-Rosen potential \ defined in the half-space $\xi >\xi _{0}$. As a
direct path integration is not possible, we can construct the corresponding
Green's function in terms of the Green's function in the interval $%
\mathbb{R}
^{+}$ with the help of the perturbation expansion method discussed in detail
in the literature \cite{Benamira1,Benamira2,Grosche} and it is not necessary
to repeat here. We find, all calculations done, that%
\begin{equation}
G^{0<\left\vert q\right\vert <1}(\xi ^{\prime \prime },\xi ^{\prime },%
\widetilde{E}_{0})=G_{MR}(\xi ^{\prime \prime },\xi ^{\prime };\widetilde{E}%
_{0})-\frac{G_{MR}(\xi ^{\prime \prime },\xi _{0};\widetilde{E}%
_{0})G_{MR}(\xi _{0},\xi ^{\prime };\widetilde{E}_{0})}{G_{MR}(\xi _{0},\xi
_{0};\widetilde{E}_{0})},  \label{F43}
\end{equation}%
where $G_{MR}(\xi ^{\prime \prime },\xi ^{\prime };\widetilde{E}_{0})$ is
the Green's function (\ref{F21}), for $l=0$.

The energy spectrum is determined by the poles of the Green's function (\ref%
{F43}), i.e. by the equation $G_{MR}(\xi _{0},\xi _{0};\widetilde{E}_{0})=0$%
, or as well by the transcendental equation%
\begin{equation}
_{2}F_{1}\left(
M_{1}-L_{E_{n_{r}}},L_{E_{n_{r}}}+M_{1}+1,M_{1}-M_{2}+1;\left\vert
q\right\vert \right) =0,  \label{F44}
\end{equation}%
where

\begin{equation}
\left\{ 
\begin{array}{c}
L_{E_{n_{r}}}=-\frac{1}{2}+\frac{1}{2\hbar \alpha }\sqrt{2M\left( \frac{%
D_{e}e^{4\alpha r_{e}}}{\left\vert q\right\vert ^{2}}-E_{n_{r}}\right) }, \\ 
M_{1}=\delta _{0}+\frac{1}{2\hbar \alpha }\sqrt{2M\left(
D_{e}-E_{n_{r}}\right) }, \\ 
M_{2}=\delta _{0}-\frac{1}{2\hbar \alpha }\sqrt{2M\left(
D_{e}-E_{n_{r}}\right) }, \\ 
\delta _{0}=\frac{1}{2}\sqrt{1+\frac{8MU_{2}}{\hbar ^{2}\alpha ^{2}}}.%
\end{array}%
\right.  \label{F45}
\end{equation}%
The transcendental equation (\ref{F44}) can be solved numerically to
determine the energy levels. The corresponding reduced radial wave functions
are of the form:%
\begin{eqnarray}
\chi _{n_{r}}^{0<\left\vert q\right\vert <1}(r) &=&C\left( 1-\left\vert
q\right\vert e^{-2\alpha r}\right) ^{\delta _{0}+\frac{1}{2}}\left(
\left\vert q\right\vert e^{-2\alpha r}\right) ^{\frac{1}{2\hbar \alpha }%
\sqrt{2M\left( D_{e}-E_{n_{r}}\right) }}  \notag \\
&&\times \text{ }_{2}F_{1}\left(
M_{1}-L_{E_{n_{r}}},L_{E_{n_{r}}}+M_{1}+1,M_{1}-M_{2}+1;\left\vert
q\right\vert e^{-2\alpha r}\right) ,  \notag \\
&&  \label{F46}
\end{eqnarray}%
where $C$ is a constant factor.

\section{Deformed modified Rosen-Morse potential}

For $q>0$, the improved Tietz potential is analogous to the deformed
modified Rosen-Morse potential

\begin{equation}
V_{q}(r)=V_{0}+V_{1}\tanh _{q}\left( \alpha r\right) +V_{2}\tanh
_{q}^{2}\left( \alpha r\right) .  \label{F47}
\end{equation}%
The constants $V_{0},$ $V_{1}$ and $V_{2}$ are given by 
\begin{equation}
\left\{ 
\begin{array}{c}
V_{0}=\frac{D_{e}}{4}\left( \frac{e^{2\alpha r_{e}}}{_{q}}-1\right) ^{2}, \\ 
V_{1}=\frac{D_{e}}{2}\left( 1+\frac{e^{2\alpha r_{e}}}{q}\right) \left( 1-%
\frac{e^{2\alpha r_{e}}}{q}\right) , \\ 
V_{2}=\frac{D_{e}}{4}\left( \frac{e^{2\alpha r_{e}}}{q}+1\right) ^{2}.%
\end{array}%
\right.  \label{F48}
\end{equation}

As the approximation to the centrifugal potential term adopted above does
not apply to the case of the Rosen-Morse potential (see, for example, the
discussion of the validity of an approximation of this type in our previous
work\cite{Khodja}), we will be satisfied with the study of the $s-$waves $%
(l=0)$ by evaluating the Green's function in order to find the energy levels
and the corresponding wave functions of the bound states. In this case, the
radial Green's function (\ref{F3}) is written%
\begin{equation}
G_{0}^{q>0}(r^{\prime \prime },r^{\prime };E)=\int_{0}^{\infty }dT\exp
\left( \frac{i}{\hbar }ET\right) K_{0}^{q>0}\left( r^{\prime \prime
},r^{\prime };T\right) ,  \label{F49}
\end{equation}%
where the propagator $K_{0}^{q>0}\left( r^{\prime \prime },r^{\prime
};T\right) $ is given by%
\begin{equation}
K_{0}^{q>0}\left( r^{\prime \prime },r^{\prime };T\right) =\int \mathit{D}%
r(t)\exp \left\{ \frac{i}{\hbar }\int_{0}^{T}\left( \frac{M}{2}\overset{.}{r}%
^{2}-V_{q}(r)\right) dt\right\} .  \label{F50}
\end{equation}%
To evaluate the Green's function (\ref{F49}), we perform the coordinate- and
time transformations $u=\alpha r-\frac{1}{2}\ln q,\left( u\in \left]
u_{0},\infty \right[ \text{, }u_{0}=-\frac{1}{2}\ln q\right) $ and $\frac{dt%
}{ds}=\frac{1}{\alpha ^{2}}$. As a result of these transformations, the
Green's function (\ref{F49}) takes the form%
\begin{eqnarray}
G_{0}^{q>0}(r^{\prime \prime },r^{\prime };E) &=&\frac{1}{\alpha }%
G_{0}^{q>0}(u^{\prime \prime },u^{\prime };\mathcal{E}_{0})  \notag \\
&=&\frac{1}{\alpha }\int_{0}^{\infty }dS\exp \left( \frac{i}{\hbar }\mathcal{%
E}_{0}S\right) P_{0}^{q>0}(u^{\prime \prime },u^{\prime };S),  \label{F51}
\end{eqnarray}%
where%
\begin{equation}
\mathcal{E}_{0}=\frac{1}{\alpha ^{2}}\left[ E-\left( V_{0}+V_{2}\right) %
\right] ,  \label{F52}
\end{equation}%
and\ \ 
\begin{equation}
P_{0}^{q>0}(u^{\prime \prime },u^{\prime };S)=\int \mathit{D}u(s)\exp
\left\{ \frac{i}{\hbar }\int_{0}^{S}\left( \frac{M}{2}\overset{.}{u}%
^{2}-V(u)\right) ds\right\} ,  \label{F53}
\end{equation}%
with 
\begin{equation}
V(u)=\frac{V_{1}}{\alpha ^{2}}\tanh u-\frac{V_{2}}{\alpha ^{2}\cosh ^{2}u};%
\text{ \ }u>u_{0}.  \label{F54}
\end{equation}%
Due to the fact that $V(u)$ is the Rosen-Morse potential defined in the
half-space $u>u_{0}$, the Green's function (\ref{F51}) can be constructed in
terms of the Green's function in the entire $%
\mathbb{R}
$, by a method similar to that used in the literature \cite%
{Benamira1,Benamira2,Grosche}. Hence the solution of (\ref{F51}) is easily
found to be

\begin{equation}
G_{0}^{q>0}(u^{\prime \prime },u^{\prime };\mathcal{E}_{0})=G_{RM}(u^{\prime
\prime },u^{\prime };\mathcal{E}_{0})-\frac{G_{RM}(u^{\prime \prime },u_{0};%
\mathcal{E}_{0})G_{RM}(u_{0},u^{\prime };\mathcal{E}_{0})}{%
G_{RM}(u_{0},u_{0};\mathcal{E}_{0})},  \label{F55}
\end{equation}%
with the Green's function $G_{RM}(u^{\prime \prime },u^{\prime };\mathcal{E}%
_{0})$ given by

\begin{eqnarray}
G_{RM}(u^{\prime \prime },u^{\prime };\mathcal{E}_{0}) &=&-\frac{iM}{\hbar }%
\frac{\Gamma \left( M_{1}-L_{\mathcal{E}_{0}}\right) \Gamma \left( L_{%
\mathcal{E}_{0}}+M_{1}+1\right) }{\Gamma \left( M_{1}+M_{2}+1\right) \Gamma
(M_{1}-M_{2}+1)}  \notag \\
&&\times \left( \frac{1-\tanh u^{\prime }}{2}\frac{1-\tanh u^{^{\prime
\prime }}}{2}\right) ^{\frac{M_{1}+M_{2}}{2}}  \notag \\
&&\times \left( \frac{1+\tanh u^{\prime }}{2}\frac{1+\tanh u^{^{\prime
\prime }}}{2}\right) ^{\frac{M_{1}-M_{2}}{2}}  \notag \\
&&\times \text{ }_{2}F_{1}\left( M_{1}-L_{\mathcal{E}_{0}},L_{\mathcal{E}%
_{0}}+M_{1}+1,M_{1}-M_{2}+1;\frac{1+\tanh u_{<}}{2}\right)  \notag \\
&&\times \text{ }_{2}F_{1}\left( M_{1}-L_{\mathcal{E}_{0}},L_{\mathcal{E}%
_{0}}+M_{1}+1,M_{1}+M_{2}+1;\frac{1-\tanh u_{>}}{2}\right) .  \notag \\
&&  \label{F56}
\end{eqnarray}%
Here we have used the abbreviations%
\begin{equation}
\left\{ 
\begin{array}{c}
L_{\mathcal{E}_{0}}=-\frac{1}{2}+\sqrt{\frac{1}{4}+\frac{2MV_{2}}{\hbar
^{2}\alpha ^{2}}}, \\ 
M_{1}=\frac{1}{2\hbar \alpha }\left( \sqrt{2M\left(
V_{0}+V_{1}+V_{2}-E\right) }+\sqrt{2M\left( V_{0}-V_{1}+V_{2}-E\right) }%
\right) , \\ 
M_{2}=\frac{1}{2\hbar \alpha }\left( \sqrt{2M\left(
V_{0}+V_{1}+V_{2}-E\right) }-\sqrt{2M\left( V_{0}-V_{1}+V_{2}-E\right) }%
\right) .%
\end{array}%
\right.  \label{F57}
\end{equation}

The bound state energy levels $E_{n_{r}}$ can be obtained from the poles of
the radial Green's function (\ref{F55}) and are determined by the
transcendental equation%
\begin{eqnarray}
_{2}F_{1}\left( M_{1}-L_{\mathcal{E}_{0}},L_{\mathcal{E}%
_{0}}+M_{1}+1,M_{1}+M_{2}+1;\frac{q}{1+q}\right) &=&0;\text{ \ \ \ for }%
E=E_{n_{r}},  \notag \\
&&  \label{F58}
\end{eqnarray}%
which can be solved numerically. The reduced radial wave functions with $%
E=E_{n_{r}}$ are readily obtained to be

\begin{eqnarray}
\chi _{n_{r}}^{q>0}(r) &=&C\left( \frac{q}{e^{2\alpha r}+q}\right) ^{\frac{1%
}{2\hbar \alpha }\sqrt{2M\left( D_{e}-E\right) }}\left( \frac{1}{%
1+qe^{-2\alpha r}}\right) ^{\frac{1}{2\hbar \alpha }\sqrt{2M\left( \frac{%
D_{e}}{q^{2}}e^{4\alpha r_{e}}-E\right) }}  \notag \\
&&\times \text{ }_{2}F_{1}\left( M_{1}-L_{\mathcal{E}_{0}},L_{\mathcal{E}%
_{0}}+M_{1}+1,M_{1}+M_{2}+1;\frac{q}{e^{2\alpha r}+q}\right) ,  \notag \\
&&  \label{F59}
\end{eqnarray}%
where $C$ is a constant factor.

\section{Morse potential}

When $q=0$, the Tietz potential given by Eq. (\ref{F1b}) turns to the Morse
potential 
\begin{equation}
V_{M}(r)=D_{e}\left( 1-e^{-2\alpha \left( r-r_{e}\right) }\right) ^{2}.\text{%
\ }  \label{F60}
\end{equation}

In this case, it can be seen from Eqs. (\ref{F57}) that

\bigskip

\begin{equation}
\left\{ 
\begin{array}{c}
L_{\mathcal{E}_{0}}\underset{q\rightarrow 0}{\simeq }-\frac{1}{2}+\lambda
\left( 1+\frac{e^{2\alpha r_{e}}}{q}\right) , \\ 
M_{1}\underset{q\rightarrow 0}{\simeq }\mu +\lambda \frac{e^{2\alpha r_{e}}}{%
q}, \\ 
M_{2}\underset{q\rightarrow 0}{\simeq }\mu -\lambda \frac{e^{2\alpha r_{e}}}{%
q},%
\end{array}%
\right.  \label{F61}
\end{equation}%
where we have identified%
\begin{equation}
\text{\ \ \ \ }\mu =\frac{\sqrt{2M\left( D_{e}-E\right) }}{2\hbar \alpha },
\label{F62}
\end{equation}%
and 
\begin{equation}
\lambda =\frac{\sqrt{2MD_{e}}}{2\hbar \alpha }.\text{ \ }  \label{F63}
\end{equation}

Using the Gauss's transformation formula (\ref{F30}) together with the
property of the hypergeometric function \cite{Landau}

\begin{equation}
\underset{\beta \rightarrow \infty }{\lim }\text{ }_{2}F_{1}(\alpha ,\beta
,\gamma ;\frac{z}{\beta })=\text{ }_{1}F_{1}(\alpha ,\gamma ;z),  \label{a64}
\end{equation}%
the relation between the confluent hypergeometric function and the Whittaker
functions (see Ref. \cite{Gradshtein}, p. 1059, Eqs. (9.220.2) and (9.220.3))

\begin{equation}
M_{\lambda ,\mu }(z)=z^{\mu +\frac{1}{2}}e^{-\frac{z}{2}}\text{ }%
_{1}F_{1}\left( \frac{1}{2}-\lambda +\mu ,1+2\mu ;z\right) ,  \label{F65}
\end{equation}%
\begin{equation}
M_{\lambda ,-\mu }(z)=z^{-\mu +\frac{1}{2}}e^{-\frac{z}{2}}\text{ }%
_{1}F_{1}\left( \frac{1}{2}-\lambda -\mu ,1-2\mu ;z\right) ,  \label{F66}
\end{equation}%
and the formula (see Ref. \cite{Gradshtein}, p. 1059, Eq. (9.220.4))

\begin{equation}
W_{\lambda ,\mu }(z)=\frac{\Gamma \left( -2\mu \right) }{\Gamma \left( \frac{%
1}{2}-\lambda -\mu \right) }M_{\lambda ,\mu }(z)+\frac{\Gamma \left( 2\mu
\right) }{\Gamma \left( \frac{1}{2}-\lambda +\mu \right) }M_{\lambda ,-\mu
}(z),  \label{F67}
\end{equation}%
it can be shown that, after some simple calculation, the Green's function (%
\ref{F55}) reduces to the well known Green's function associated with the
radial Morse potential%
\begin{equation}
G_{M}\left( r^{\prime \prime },r^{\prime };E_{0}\right) =-\frac{iM}{\hbar }%
\frac{\Gamma \left( \frac{1}{2}-\lambda +\mu \right) }{\Gamma \left( 2\mu
+1\right) \sqrt{z^{\prime \prime }z^{\prime }}}e^{\frac{z^{\prime \prime
}+z^{\prime }}{2}}M_{\lambda ,\mu }(z^{\prime })W_{\lambda ,\mu }(z^{\prime
\prime });\text{ \ \ }z^{\prime \prime }>z^{\prime },  \label{F68}
\end{equation}%
where $z=2\lambda e^{-2\alpha \left( r-r_{e}\right) }$.

From (\ref{F58}), it follows that

\begin{eqnarray}
&&\underset{q\rightarrow 0}{\lim }\text{ }_{2}F_{1}\left( M_{1}-L_{\mathcal{E%
}_{0}},L_{\mathcal{E}_{0}}+M_{1}+1,M_{1}+M_{2}+1;\frac{q}{1+q}\right)  \notag
\\
&=&\text{ }_{1}F_{1}\left( \frac{1}{2}-\lambda +\mu ,2\mu +1;2\lambda
e^{2\alpha r_{e}}\right) =0.  \label{F69}
\end{eqnarray}%
Following Fl\"{u}gge \cite{Flugge}, this transcendental equation can be
solved approximatively. Since, in general, the values of $\lambda e^{2\alpha
r_{e}}\gg 1$ for the standard diatomic molecules, the asymptotic behavior of
the confluent hypergeometric function (\ref{F69}) can be taken into
consideration to show that

\begin{equation}
\frac{1}{2}-\lambda +\mu =-n_{r},\text{ }n_{r}=0,1,2,....  \label{F70}
\end{equation}%
Upon inserting the values of $\lambda $ and $\mu $ in (\ref{F70})$,$ one
finds the well-known energy levels for the radial Morse potential:%
\begin{equation}
E_{n_{r}}=-\frac{2\hbar ^{2}\alpha ^{2}}{M}\left[ \left( n_{r}+\frac{1}{2}%
\right) ^{2}-\left( n_{r}+\frac{1}{2}\right) \frac{\sqrt{2MD_{e}}}{\hbar
\alpha }\right] ,  \label{F71}
\end{equation}%
with $n_{r\max }=\left\{ \frac{\sqrt{2MD_{e}}}{2\hbar \alpha }-\frac{1}{2}%
\right\} $.

The corresponding wave functions are found, in the limiting case $q$ $%
\rightarrow $ $0$, from (\ref{F59}) to be

\begin{eqnarray}
\chi _{n_{r}}^{q=0}(r) &=&\mathcal{N}\exp \left[ -\frac{\sqrt{2MD_{e}}}{%
2\hbar \alpha }e^{-2\alpha \left( r-r_{e}\right) }\right] \left( e^{-2\alpha
\left( r-r_{e}\right) }\right) ^{\frac{\sqrt{2MD_{e}}}{2\hbar \alpha }-n_{r}-%
\frac{1}{2}}  \notag \\
&&\times \text{ }_{1}F_{1}\left( -n_{r},2n_{r}-\frac{\sqrt{2MD_{e}}}{\hbar
\alpha }+2;\frac{\sqrt{2MD_{e}}}{\hbar \alpha }e^{-2\alpha (r-r_{e})}\right)
,  \notag \\
&&  \label{F72}
\end{eqnarray}%
where $\mathcal{N}$\ is the normalization constant.

\section{Discussions}

Firstly, it should be noted that there are several special cases of improved
Tietz potential that have been analyzed through different methods. For $q=1$%
, the deformed modified Rosen-Morse potential (49) becomes the general
radial Rosen-Morse potential%
\begin{equation}
V(r)=V_{0}+V_{2}+V_{1}\tanh \left( \alpha r\right) -\frac{V_{2}}{\cosh
^{2}\left( \alpha r\right) }  \label{F73}
\end{equation}%
which is similar to that of the Natanzon potential. The results obtained by
Natanzon in Ref. \cite{Natanzon} using a transformation of the Schr\"{o}%
dinger equation into a hypergeometric equation and by Wu et al. \cite{Wu} in
a work based on the group theory approach are not satisfactory. When $%
V_{1}=0,$ and $V_{0}=-V_{2},$ the potential (\ref{F73}) reduces to the
radial Rosen-Morse potential (also called radial modified P\"{o}schl-Teller
potential) for which the Schr\"{o}dinger equation has been solved by Nieto
in Ref. \cite{Nieto}, for the $s-$waves from that with the symmetric
Rosen-Morse potential by imposing that the general wave function must have
continuous derivatives at the origin to keep only the odd solutions. Given
the results obtained, it is concluded that this is a bad idea.

Secondly, to obtain closed-form expressions for anharmonicity constants $%
\omega _{e}x_{e}$ and $\omega _{e}y_{e}$ in terms of the energy spectrum
expression, Sun et al. \cite{Sun} have established the equivalence, for the
three-dimensional case, between the deformed modified Rosen-Morse potential
and the Tietz potential. However, they have inappropriately adapted the
expression of the vibrational energy levels associated with the
one-dimensional deformed modified potential \cite{Egrifes} to that of the
Tietz potential.

Finally, It is well known that the supersymmetry approach in quantum
mechanics is based on the factorization method that applies to problems for
which wave equations admit orthogonal polynomials as solutions. Therefore,
the problem solution of a new deformed Schi\"{o}berg-type potential is
partially correct $(q<-1)$ via this method contrary to what is claimed by
Mustafa \cite{Mustafa}. For $-1<q<0$ or $q>0$, we have a problem with
Dirichlet boundary conditions. In this case, the solutions of the Schr\"{o}%
dinger equation are hypergeometric series $_{2}F_{1}\left( a,b,c,z\right) $
with transcendental equations for energy levels. The same potential has been
treated within the path integral approach by Amrouche et al. \cite{Amrouche}
without incorporating the Dirichlet boundary conditions into the path
integral. Likewise, their solutions for the cases $-1<q<0$ and $q>0$ must be
discarded as that of Mustafa.

\section{Conclusion}

In the above, we have presented a method to solve completely an improved
form of the Tietz potential in terms of the path integral formalism. As we
have shown, the path integral for the Green's function associated with this
potential can not be constructed for any deformation parameter $q$ in a
unified way because the parameter $q$ characterizes various shapes of the
potential. This potential has a strong singularity at $r=r_{0}=\frac{1}{%
2\alpha }\ln \left\vert q\right\vert $ when $q<0$ and the limiting case $%
q=0. $ Contrary to what has been reported in the literature, the quantum
treatment of the problem with this potential by any method requires to
consider three cases separately. When $\left\vert q\right\vert \geq 1$ and $%
\frac{1}{2\alpha }\ln \left\vert q\right\vert <r<+\infty $, by adopting a
suitable approximation for the centrifugal potential term and formulating
the path integral in terms of the standard Manning-Rosen potential, the
radial Green's function for any $l$ wave is stated in a closed form, from
which the energy spectrum and the normalized wave functions are extracted.
For $\left\vert q\right\vert <1$, the path integral for the Green function
associated with the $s$ waves is also formulated in terms of the
Manning-Rosen potential defined in the half-space $\xi >-\frac{1}{2\alpha }%
\ln \left\vert q\right\vert $ and when $q>0$, it is expressed in terms of
the Rosen-Morse potential in the half-space $u>-\frac{1}{2\alpha }\ln q$. In
both cases, we have shown that the $s$ state energy levels are determined by
a transcendental equation involving the hypergeometric function. Naturally,
in the limit $q\rightarrow 0$, our results can be reduced to those for the
radial Morse potential. $\ $

\end{document}